\documentclass{article}

\usepackage[numbers, compress]{natbib}
\usepackage{tabularx}
\usepackage{colortbl}
\usepackage[verbose=true,letterpaper]{geometry}
\usepackage[utf8]{inputenc} 
\usepackage[T1]{fontenc}    
\usepackage{hyperref}       
\usepackage{url}            
\usepackage{booktabs}       
\usepackage{amsfonts}       
\usepackage{nicefrac}       
\usepackage{microtype}      

\usepackage{amsmath}
\usepackage[plain]{fancyref}
\usepackage{graphicx}
\usepackage[textsize=small, disable]{todonotes}
\usepackage{listings}
\newcommand{\todoref}[1]{\todo[linecolor=blue,backgroundcolor=blue!25,bordercolor=blue]{add reference}}
\renewcommand{\vec}[1]{{\bf #1}}

\title{Closing the loop between neural network simulators and the OpenAI Gym}

\newcommand{\inm}{Institute of Neuroscience and Medicine (INM-6)}
\newcommand{\ias}{Institute for Advanced Simulation (IAS-6)}
\newcommand{\jbi}{JARA-Institute Brain Structure Function Relationship (JBI 1 / INM-10)}

\newcommand{\rcj}{
  Research Centre J\"ulich, J\"ulich, Germany
}

\newcommand{\figuredirnestrl}{.}

\author{
  Jakob Jordan$^{1,2,3}$\thanks{These authors contributed equally to this study.}\;\thanks{Corresponding author: \texttt{j.jordan@fz-juelich.de}} \and Philipp Weidel$^{2,1,3*}$ \and Abigail Morrison$^{2,1,3}$
}

\date{
  \small
  $^1$ \inm, \rcj \\
  $^2$ \ias, \rcj \\
  $^3$ \jbi, \rcj
}

\begin{document}

\maketitle

\begin{abstract}
Since the enormous breakthroughs in machine learning over the last decade, functional neural network models are of growing interest for many researchers in the field of computational neuroscience.
One major branch of research is concerned with biologically plausible implementations of reinforcement learning, with a variety of different models developed over the recent years.
However, most studies in this area are conducted with custom simulation scripts and manually implemented tasks.
This makes it hard for other researchers to reproduce and build upon previous work and nearly impossible to compare the performance of different learning architectures.
In this work, we present a novel approach to solve this problem, connecting benchmark tools from the field of machine learning and state-of-the-art neural network simulators from computational neuroscience.
This toolchain enables researchers in both fields to make use of well-tested high-performance simulation software supporting biologically plausible neuron, synapse and network models and allows them to evaluate and compare their approach on the basis of standardized environments of varying complexity.
We demonstrate the functionality of the toolchain by implementing a neuronal actor-critic architecture for reinforcement learning in the NEST simulator and successfully training it on two different environments from the OpenAI Gym.

  
\end{abstract}

\section{Introduction}
The last decade has witnessed major progress in the field of machine learning, moving from small-scale toy problems to large-scale real-world applications including image \cite{Krizhevsky12_1097} and speech recognition \cite{Hinton12_82}, complex motor-control tasks \cite{Mnih2016_1928} and playing (video) games at super-human performance \cite{Mnih15_529,Silver16_484}.
This progress has been driven mainly by an increase in computing power, especially by training deep networks on graphics processing units \cite{Raina09_873}, and conceptual breakthroughs like layer-wise pretraining \cite{Hinton2006_504,Bengio07_253} or dropout \cite{Hinton2012_1207}.
Even so, this rate of progress would not have been possible without the wide availability of high-performance ready-to-use tools, e.g., Torch \cite{Collobert2002}, Theano \cite{James2010}, Caffe \cite{Jia2014_1408}, TensorFlow \cite{Abadi2016_1603} and standardized benchmarks and learning environments, such as the MNIST \cite{Lecun1998}, CIFAR \cite{Krizhevsky09} and ImageNET \cite{Deng09} datasets, and the MuJoCo \cite{Todorov2012_5026}, ALE \cite{Bellemare2015} and OpenAI Gym \cite{Brockman2016} toolkits.
While ready-to-use tools allow researchers to focus on important aspects rather than basic implementation details, standardized benchmarks can guide the community as a whole towards promising approaches, as for example in the case of convolutional networks through the ImageNET competition \cite{Russakovsky15_211}.

Similarly, researchers in the field of computational neuroscience have benefited from the increase of computational power and achieved many conceptual breakthroughs over the last decade, with a plethora of new neuron, synapse and network models being developed.
Thanks to a variety of software projects, researchers have access to simulators for all scales of neural systems from molecular simulations \cite{Wils09_15} over complex neuron \cite{Carnevale06,Bower07_1383} and network models \cite{Gewaltig_07_11204,Goodman09_192,Bekolay2013_7} to whole brain simulations using neural fields \cite{SanzLeon13}.
However, in computational neuroscience no generally accepted set of benchmarks exist so far (but see \cite{Gerstner09_379}).
While it is desirable to compare different neural network models with respect to biological plausibility, explanatory power and functional performance, only the latter lends itself to the definition of quantitative benchmarks.

One particular area in which the lack of standardized benchmarks is apparent is research into reinforcement learning (RL) in neurobiological substrates.
Inspired by behavioural experiments, RL is concerned with the ability of organisms to learn from previous experiences to optimize their behavior in order to maximize reward and avoid punishment (see, e.g., \cite{Sutton98}).
RL has a long tradition in the field of machine learning which has led to several powerful algorithms, such as SARSA and Q-learning \cite{Watkins1989}.
Similarly, a large variety of neurobiological models have been proposed in recent years \cite{Vasilaki09,Izhikevich07_2443,Urbanczik09_250,Potjans09_301,Fremaux10_13326,Jitsev12,Fremaux2013_e1003024,Rasmussen2014,Friedrich2014_1450002,Rombouts2015_e1004060,Baladron2015_1,Aswolinskiy2015_36,Friedrich2016_1529,Rueckert2016_6}.
However, only a small proportion of these rely on publicly available simulators and all of them employ custom built environments.
Even for fairly simple environments, this has led to a situation where different network models are difficult to compare and reproduce, thus creating a fragmentation of research efforts.
Instead of building upon and extending existing models, researchers are forced to spend too much time on recreating basic methods for custom implementations.

This issue has led to the Human Brain Project (HBP) \cite{HumanBrainProject14} dedicating significant resources of a subproject (SP10, Neurorobotics) to the development of the necessary infrastructure that allows users to conduct robotic experiments in virtual environments and connect these to their neural network implementations with an easy to use web interface.
This approach however, specifically addresses the need of researchers developing neuronal controllers for robotic applications.

In contrast, the OpenAI Gym \cite{Brockman2016} provides a rich and generic collection of standardized RL environments developed to support the machine learning community in evaluating and comparing algorithms.
All environments are accessible via a simple, unified interface, that requires an agent to supply an action and returns an observation and reward for its current state.
The toolkit includes a range of different environments with varying levels of complexity ranging from low-dimensional fully discrete (e.g., \emph{FrozenLake}) to high-dimensional fully continuous tasks (e.g., \emph{Humanoid}).
Consistency of the OpenAI Gym environments across different releases supports researchers in reproduction and extension of previous work and allows systematic benchmarking and comparison of learning algorithms and their implementations.
The easy accessibility of different tasks fosters progress by allowing researchers to focus on learning algorithms instead of basic implementation details of particular environments and provokes researchers to evaluate the performance of their algorithms on many different tasks.

To make this comprehensive resource available to the computational neuroscience community, we developed a toolchain to interface neural network simulators with the OpenAI Gym.
Using this toolchain, researchers can rely on well-tested, high-performance simulation engines to power their models and evaluate them against a curated set of standardized environments, allowing more time to focus on neurobiological questions.

In the next section we introduce additional pre-existing components on which our toolchain relies, and afterwards discuss how it links the different tools. We demonstrate its functionality by implementing a neural actor-critic in NEST (\cite{Gewaltig_07_11204}, NEural Simulation Tool) and successfully training it on two different environments from the OpenAI Gym.

\begin{figure}
\center
\includegraphics[width=0.8\textwidth]{\figuredirnestrl/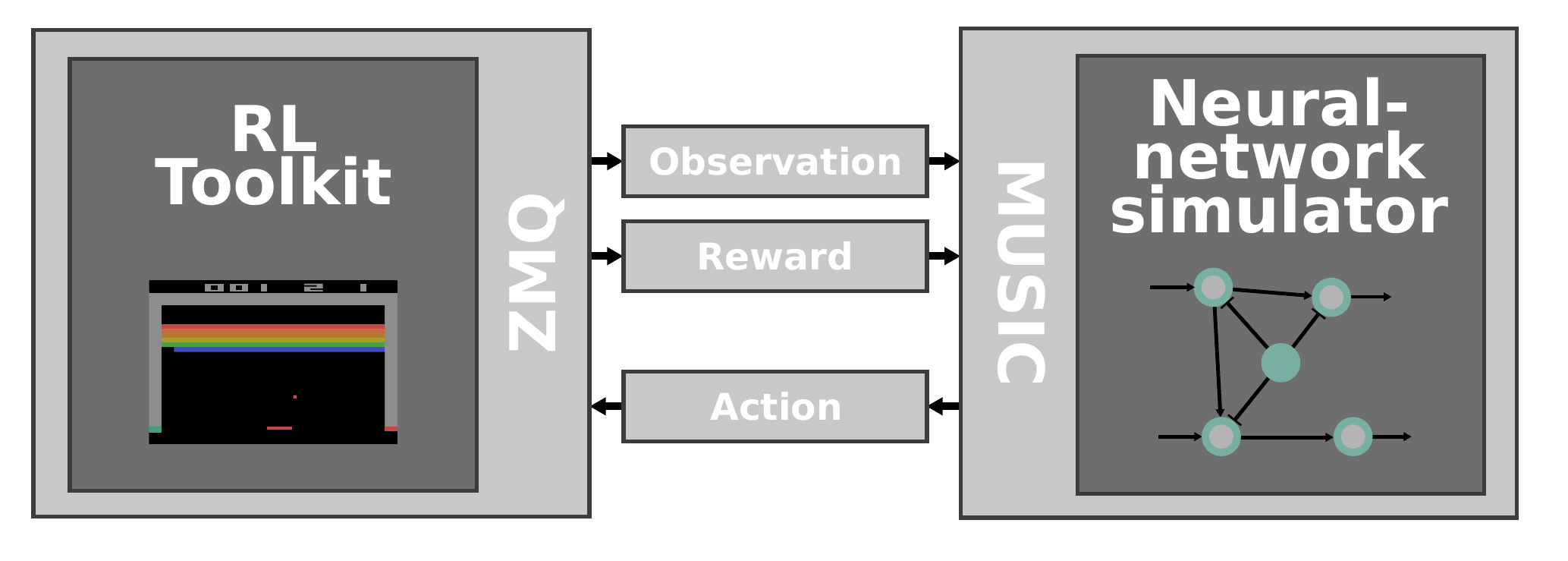}
\caption{{\bf Interfacing RL toolkits with neural network simulators.} The RL toolkit (left) is responsible for emulating an environment that provides observations and rewards which are communicated via ZeroMQ sockets and MUSIC adapters (middle) to a neural network simulator (right). Concurrently, the activity of the simulated neural network is transformed to an action and fed back to the RL toolkit.}
\label{fig:general_interface}
\end{figure}


\section{Pre-existing components}
NEST is a neural simulator designed for the efficient simulation of large-scale networks of simple neuron models with biophysically realistic connectivity.
The simulation kernel scales from small simulations on a laptops to super computers, with the largest simulation containing about $10^9$ neurons and $10^{13}$ synapses, corresponding to about 10\% of the human cortex at the resolution of individual cells and connections \cite{Kunkel14_78}. 
NEST is actively developed and maintained by the NEST initiative in collaboration with the community and freely available under the GPLv2 and is supported by the HBP with the explicit aim of widespread long-term availability and maintainability.
%

While the network implementation that we present in the results section relies on the NEST simulator, the toolchain can also be used with other simulators that support the MUSIC library, for example NEURON \cite{Carnevale06}.
The MUlti-SImulation Coordinator is a multi-purpose middleware for neural network simulators built on top of MPI (Message Passing Interface) that enables online interaction of different simulation engines \cite{Djurfeldt10}.
MUSIC provides named MPI channels, referred to as MUSIC ports, which allow the user to set up communication streams between several processes.
%
While originally intented to distribute a single neural network model across different simulators, the MUSIC library can also be used to connect neural simulators with robotic applications.

For this purpose the ROS-MUSIC Toolchain (RMT) \cite{Weidel16_10} was recently developed, providing an interface from MUSIC to the Robotic Operating System (ROS) \cite{Quigley09_5}.
ROS is the most popular middleware in the robotic community which can interact with many robotic simulators and hardware platforms.
The RMT allows exchange of well-defined messages between ROS and MUSIC via stand-alone executables, so called adapters, that were designed with a focus on modularity.
The toolchain contains several different adapters each performing a rather simple operation on streams of inputs (e.g., filtering).
By concatinating several adapters, the overall transformation of the original data can become more complex, for example converting high-dimensional continuous data (e.g., sensory data) to low-dimensional discrete data (e.g., action potentials) or vice-versa.
More information and introductory examples can be found on GitHub.\footnote{\href{https://github.com/incf-music/ros-music-adapters}{https://github.com/incf-music/ros-music-adapters}}


\section{Results}
To enable the online interaction of neural network simulators and the OpenAI Gym we rely on two different libraries, MUSIC, to interface with the neural simulator, and ZeroMQ \cite{Hintjens2013} to exchange messages with the environment simulated in the OpenAI Gym. In the following we describe these two parts of the toolchain and demonstrate their functionality by interfacing a neural network simulation in NEST with two different environments.

\subsection{Extending the ROS - MUSIC toolchain}
\label{sec:zmq-adapter-ros}

We extended the RMT by adding adapters that support communication via ZeroMQ following a publish-subscribe pattern.
ZeroMQ is a messaging library that allows applications to exchange messages at runtime via sockets.
Continuously developed by a large community, it offers bindings for a variety of languages including C++ and Python, and supports most operating systems.
A single communication adapter of the RMT sends (receives) data via a ZeroMQ socket and receives (sends) data via a MUSIC port.
While the adapters can handle arbitrary data, we defined a set of specialized messages in JSON format (see supplementary material) specifically designed to communicate observations, rewards and actions as discrete or continuous real-valued variables of arbitrary dimensions, as used in the OpenAI Gym.
We chose the JSON format due to its simplicity, easy serialization and broad platform support.

In addition to the ZeroMQ adapters dedicated for communication with MUSIC, we developed several further adapters that can perform specific transformations of the data.
As discussed above, environments can be defined in continuous or discrete spaces with arbitrary dimensionality. To generate the required closed-loop functionality, the observations provided by the environment must be consistently transformed to a format that can be fed into neural network simulations. Conversely, the activity of the neural network must be interpreted and transformed into valid actions which can be executed in the environment.

A standard way to address the first issue is to introduce so called {\it place cells}. Each of these cells is tuned to a preferred (multidimensional) observation, i.e., is highly active for a specific input and less active for other inputs \cite{Fremaux2013_e1003024}. The dependence of the activity of a single place cell on observations is described by its tuning curve, often chosen as a multidimensional Gaussian. To perform the transformation of observations to activity of place cells, we implemented a {\it discretize adapter} that allows users to specify the position and width of the tuning curves of an arbitrary number of place cells. While having a certain biological plausibility \cite{Moser08_69}, one disadvantage of this approach is that the number of place cells required to cover the whole observation space evenly scales exponentially in the number of dimensions of the observation. For observations with a small number of dimensions, however, this approach is very suitable.

To perform action selection, we added several adapters that can, respectively, select the most active neuron ({\it argmax adapter}), threshold the activity across neurons to create a binary vector ({\it threshold adapter}) or linearly combine the activity of neurons across many input channels ({\it linear decoder}). Depending on the type of action required (discrete/continuous) by the environment, the user can select a single one or a combination of these. See the documentation of the RMT for detailed specifications of the adapters.

In general, we followed the design principle behind the RMT and developed modular adapters. This makes each individual adapter easy to understand and enables users to quickly extend the toolchain with their own adapters. By combining several adapters, the RMT allows arbitrarily complex transformations of the data and can hence be applied to many use-cases.

\subsection{ZeroMQ wrapper for the OpenAI Gym}
\label{sec:zeromq-gym}

The second part of the toolchain is a Python wrapper around the OpenAI Gym that exposes ZeroMQ sockets for communicating actions, observations and rewards (see \fref{fig:general_interface}). An environment in the OpenAI Gym is updated in steps. In each step, an agent needs to provide an action and receives an observation and reward for its current state.
The wrapper consists of four different threads that coordinate: (i) performing steps in an environment, (ii) receiving actions via a ZeroMQ SUB socket, (iii) publishing observations via a ZeroMQ PUB socket and (iv) publishing rewards via a ZeroMQ PUB socket.
Before spawning the threads, the wrapper starts a user-specified environment and creates the necessary communication buffers. The thread coordinating the environment reads actions from the corresponding buffer, performs single steps in the environment and updates the observation and reward buffers based on the return values of the environment. Upon detecting that a single episode has ended, e.g., by an agent reaching a certain goal position, it resets the environment and allows a user-specified break before starting another episode.
The communication threads continuously send(receive) messages via ZeroMQ and read from(write to) the corresponding buffers. All threads can be run with different update intervals, for example, to slow down movement of the agent by performing steps on a coarse time grid whilst continuously receiving action choices from the neural network simulation running on a fine time grid. The user can specify a variety of parameters via a configuration file in JSON format (see supplementary material). See the documentation for detailed specifications of the wrapper.

\subsection{Applications}
To demonstrate the functionality of the toolchain, we implemented a
neural network model with actor-critic architecture in NEST and
trained it on two different environments simulated in the OpenAI Gym.
In the first task, the agent needs to learn to perform a sequence of
actions in order to reach the top of a hill in a continous
environment. The second task is a classical grid-world in which
an agent needs to learn to navigate to a goal position in a
two-dimensional discrete environment with obstacles.  We first
describe the neural network architecture and learning rule and
afterwards discuss the network's performance on the two tasks.

\subsubsection{Neural network implementation}
\label{sec:neural-network}

We consider a temporal-difference learning algorithm \cite{Sutton98} implemented as an actor-critic architecture, originally using populations of spiking neurons \cite{Fremaux2013_e1003024}.
We translated the spike-based implementation to rate neurons, mainly to simplify the implementation by avoiding issues arising from noise introduced by spiking neuron models \cite{Potjans11_e101133,Fremaux2013_e1003024}.
The neuron dynamics we considered here are determined by the following stochastic differential equation:
\begin{align}
  \tau \frac{dz_i(t)}{dt}=-z_i(t) + \mu_i + f\left(h_i(t) - \theta_i\right) + \xi_i(t)\, ,
  \label{eq:rate_model}
\end{align}
where $\tau$ is some positive time constant, $\mu_i$ a baseline activity
level, $f(\cdot)$ some (arbitrary) activation function, $h_i(t)$ a time
dependent input field, $\theta_i$ an input threshold and $\xi_i(t)$ white
noise with a certain standard deviation $\sigma_\xi$.  The input field
$h_i(t)$ is determined by the activity of other neurons according to
$h_i(t) = \sum_j w_{ij}z_j(t)$, with $w_{ij}$ denoting the strength of
the connection (weight) from neuron $j$ to neuron $i$. Here we will
exclusively consider activation functions of the form $f(x)=x$ (linear
case), and $f(x)=\Theta(x)x$ (threshold-linear case, ``relu''). These
models have recently been added to the NEST simulator.
Their dynamics are solved on a fixed time-grid
by a stochastic-exponential-Euler method with a step size determined
by the resolution of the simulation. For more details on the model
implementation see \cite{Hahne16_arxiv}.

\begin{figure}
\center
\includegraphics[height=0.3\textheight]{\figuredirnestrl/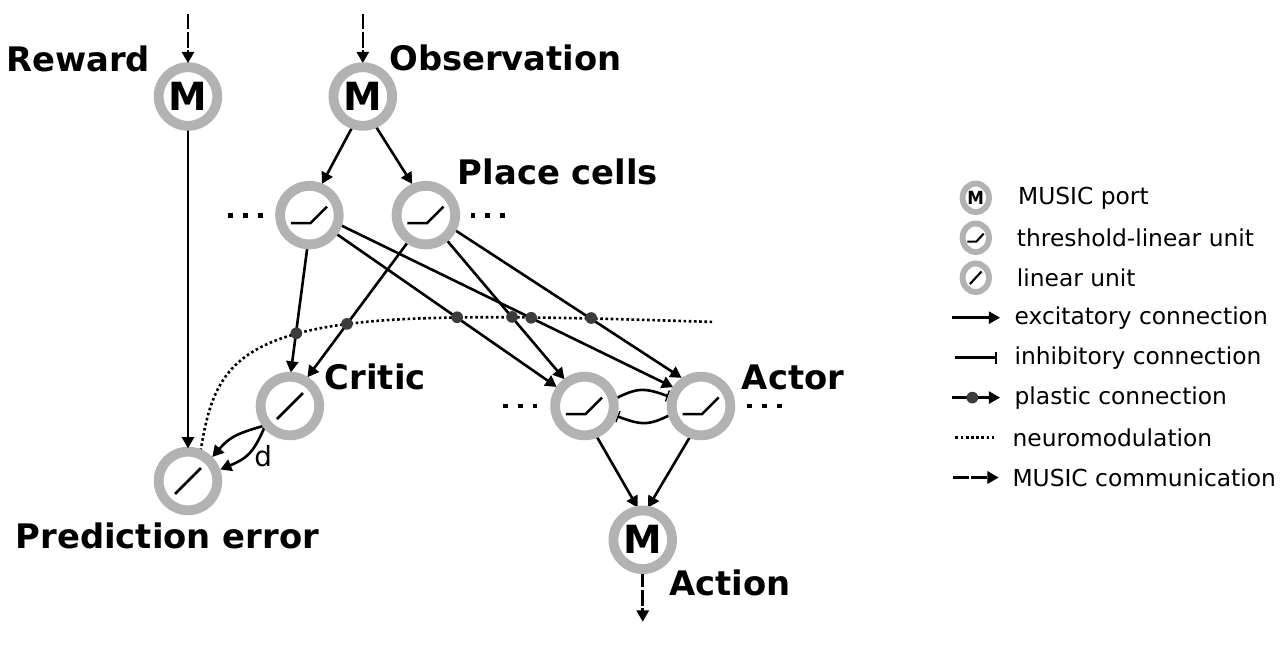}
\caption{{\bf Actor-critic architecture for RL with rate neurons}. Observations are communicated via a MUSIC input port to a population of place cells. These project on the one hand to a critic unit and on the other hand to actor units arranged in a winner-take-all circuit. The critic and an additional MUSIC input port project to a unit representing the reward-prediction error that modulates the plasticity between place cells and critic and actors, respectively. The actor units project to a MUSIC output port encoding the selected action.}
\label{fig:NN_implementation}
\end{figure}

The input layer is a population of threshold-linear rate neurons which
receive inputs through MUSIC and encode observations from the environment
(see \fref{fig:NN_implementation}). Via plastic connections these
place cells project to a single neuron representing the value that the
network assigns to the current state (the ``critic''). An additional
neuron calculates the reward-prediction error by combining the reward
received from the environment with input from the critic. Plasticity
of the projections from inputs to the critic is modulated by this
reward-prediction error (see below).

In addition, neurons in the input layer project to a population of
neurons representing the available actions (the ``actor''). To enforce
selection of a specific action, the actor units are arranged in a
winner-take-all (WTA) circuit. This is implemented by recurrent
connections between actor units that correspond to short-range
excitation and long-range inhibition, depending on the similarity of
the action that actor units encode. The activity of actor units is
transformed to a valid action and communicated to the environment via the RMT.

To derive a learning rule for the critic, we followed similar steps as in \cite{Fremaux2013_e1003024} applied to rate models (equation \eqref{eq:rate_model}). The critic activity should approximate a continous-time value function defined by \cite{Doya00}
\begin{align}
  V^\pi(\vec{s}(t)):= \int_{t'}^\infty r(\vec{s}^\pi(t'))e^{-\frac{t'-t}{\tau_r}}dt'\, .
  \label{eq:cont_value}
\end{align}
Here $\vec{s}(t)$ denotes the state of the agent at time $t$, $r(\vec{s}^\pi(t))$ denotes the reward obtained in state $\vec{s}(t)$, $\tau_r$ a discounting factor for future rewards and $\pi$ the agent's policy.
To achieve this, we define the following objective function which should be minimized by gradient descent on the weights from inputs to the critic:
\begin{align}
  E(t) = \frac{1}{2}(V^\pi(t) - z(t))^2\, ,
  \label{eq:objective_function}
\end{align}
where $z(t)$ represents the activity of the critic unit.
By performing gradient descent on equation \eqref{eq:objective_function}, using a self-consistency equation for $V^\pi(t)$ from the derivative of equation \eqref{eq:cont_value} and bootstrapping on the current prediction for the value (see supplementary material and \cite{Doya00,Fremaux2013_e1003024}), we obtain the following local Hebbian three-factor learning rule that approximately minimizes the objective function (equation \eqref{eq:objective_function}):
\begin{align}
  \Delta w_j = \eta \delta(t) x_j(t) \Theta\left(z(t) - \theta_\text{post} \right)\, ,
  \label{eq:learning_rule}
\end{align}
where $\eta$ is a learning rate, $x_j(t)$ represents the activity of the $j$th place cell, $\Theta(\cdot)$ the Heaviside function and $\theta_\text{post}$ a parameter that accounts for noise on the postsynaptic unit (see supplementary material for details). The term $\delta(t) = \dot{v}(t) + r(t) - \frac{1}{\tau_r} v(t)$ corresponds to the activity of the reward-prediction error unit, acting as a neuromodulatory signal for the Hebbian plasticity between the presynaptic ($x_j$) and postsynaptic ($z$) units. To avoid explicit calculation of the derivative, we rewrite $\delta(t)$ as:
\begin{align}
  \delta(t) \approx \left(\frac{1}{d} - \frac{1}{\tau_r} \right)v(t) - \frac{1}{d} v(t-d) + r(t)\, .
\end{align}
To approximate the derivative we hence implement two connections from the critic to the reward-prediction error unit: one instantaneous, and one with delay $d > 0$.

To learn an optimal policy, we exploit that the actor units
follow the same dynamics as the critic. Similar to
\cite{Fremaux2013_e1003024}, we hence apply the same learning rule
to the connections between the inputs and the actor units. In order
to assure that at least one actor unit is active, thus preventing a
deadlock, we introduce a minimal weight for each connection between
input and output units and add input noise to the actor units.

\subsubsection{Mountain Car}
\label{sec:mountain-car}

\begin{figure}[t]
  \center
  \includegraphics[width=1.0\textwidth]{\figuredirnestrl/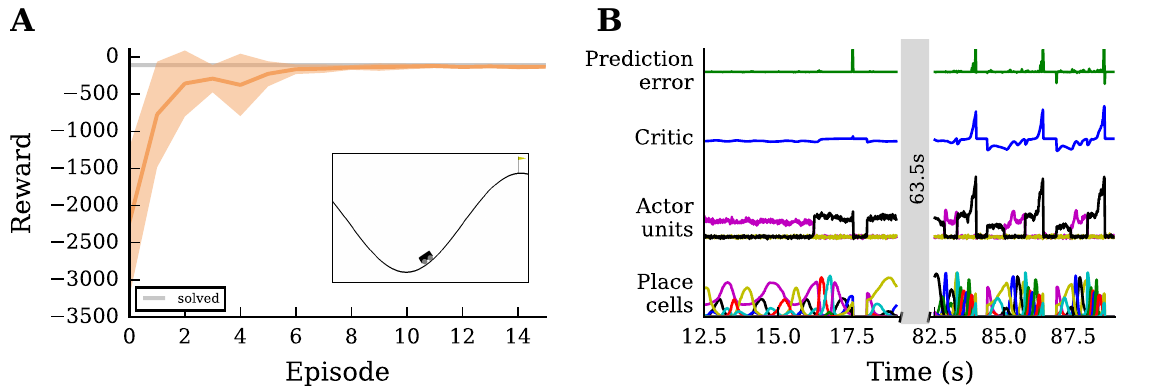}

  \caption{{\bf Network performance on an environment with continuous states and discrete actions.}
    {\bf A:} Reward obtained by the agent per episode averaged over $10$ simulations with different seeds (solid orange line). Orange band indicates $\pm$ one standard deviation. Light gray line marks average reward per episode for which the environment is considered solved. Inset: screenshot of the environment with agent (stylized vehicle), environment with valley and two hills (black solid line) and goal position (yellow flag). The agent is close to the starting position at the trough.
    {\bf B:} Activity traces of place cells (bottom), actor units (second from bottom), critic unit (second from top) and reward-prediction-error unit (top). Shown are neural activities during $6.5\;\mathrm{s}$ early (left) and late (right) in the simulation.
  }
  \label{fig:mountain_car}
\end{figure}

As an example of an environment with continous states, we consider the
\emph{MountainCar} environment. The task is to steer a toy vehicle
that starts at a valley between two hills to the top of the right one
(\fref{fig:mountain_car}{\bf A}, inset). To make the task more challenging,
the car's engine is not strong enough to reach the top in one go, so
the agent needs to learn to gain momentum by swinging back and forth
between the two hills. A single episode in this environment starts
when the agent is placed in the valley and ends when it reaches the
final position on the top of the right hill. The state of the agent is
described by two continous variables: the x-position $x(t)$ and the
x-velocity $\dot{x}(t)$. The agent can choose from three different
discrete actions that affect the velocity of the vehicle (accelerate
left, no acceleration, accelerate right). It receives
punishment from the environment in every step; the goal is to
minimize the total punishment collected over the whole episode. Since
this is difficult to implement in an actor-critic architecture
\cite{Potjans11_e101133}, we provide additional reward when the agent
reaches the final position.

To translate the agent's current state into neuronal activity, we distribute 25 place cells evenly across the two dimensional plane of possible positions and velocities using the \emph{discretize adapter} of the RMT. The actor is implemented by a WTA circuit of three units as described in \fref{sec:neural-network}. The activity of these units is transformed into an action via the \emph{argmax adapter} (\fref{sec:zmq-adapter-ros}).

Initially, the agent explores the environment by selecting random
actions. Due to the WTA circuit dynamics, a single action stays active
over an extended period of time. The constant punishment gradually
decreases the weights from the place cells to the corresponding actor unit,
eventually leading to another actor unit becoming active
(\fref{fig:mountain_car}{\bf B}, left). After a while, the agent
reaches the goal by performing actions that have not been
significantly punished. For this task the stable nature of the WTA is
advantageous, causing the agent to perform the same action repeatedly
allowing efficient exploration of the state space. After the agent has found the
goal once, the number of steps spent on exploring actions in the
following episodes is much smaller. From the sixth episode on, the
performance of the agent is already close to optimal
(\fref{fig:mountain_car}{\bf A}). After learning for about 10
episodes, the agent's performance has converged.  The value of the
final state is successfully propagated backwards over different
states, leading to a ramping of activity of the critic unit from the
start of an episode to the end (\fref{fig:mountain_car}{\bf B},
right).

Since the OpenAI Gym offers a variety of environments, we trained the same network model on an additional task with different requirements.

\subsubsection{Frozen Lake}
\label{sec:frozen-lake}

\begin{figure}[t]
  \center
  \includegraphics[width=1.0\textwidth]{\figuredirnestrl/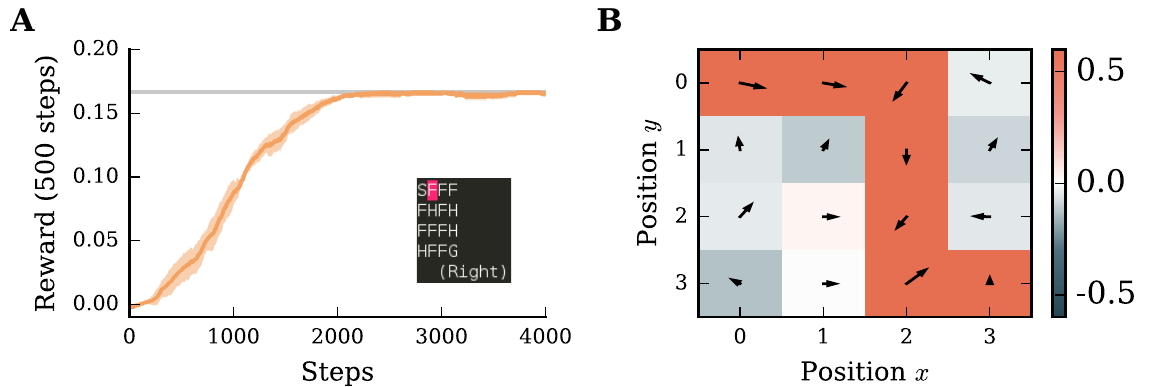}
    \caption{{\bf Network performance on a grid-world environment.}
      {\bf A}: Average reward collected by the agent over the next 500 steps (orange solid line) averaged over 5 simulations.
      Orange band indicates $\pm$ one standard deviation.
      Gray line: theoretical optimum.
      Inset: screenshot of the environment with start state (S), frozen states (F), holes (H) and goal state (G). The position of the agent is indicated in pink.
      {\bf B}: The learned policy and value map of the environment. 
      Red colors indicate positive, blue colors negative values. 
      Arrows indicate the preferred direction of movement.}
  \label{fig:grid_world}
\end{figure}

As a second application, we chose the \emph{FrozenLake} environment
consisting of a discrete set of 16 states arranged in a four-by-four
grid (\fref{fig:grid_world}{\bf A}, inset). Each state is either a start
state (S), a goal state (G), a hole (H) or a frozen state (F).
From the start position, the agent has to reach the rewarded state
by navigating over the frozen states without falling into holes which
would reset the agent to the starting position. In each step, the
agent can choose from four different actions: move west, move north,
move east and move south.  Usually, the tiles are ``slippery'', i.e.,
there is a chance that a random action is executed irrespective of the
action chosen by the agent. However, to simplify learning for
demonstration purposes, we turned this feature off. Upon reaching the
goal, the agent receives a reward of magnitude one. Since the optimal
path involves six steps from start to goal, the theoretical optimal
reward per step is $\sim0.16$.  To encourage exploration, the agent
receives a small punishment in each state, and additionally,
to speed up learning, the agent is punished for falling into holes.

Unlike in the continuous \emph{MountainCar} environment, the tuning curves of
place cells do not overlap in the discrete case, leading to sharp
transitions in the network activity. This leads to severe issues for
associating values and actions with the respective states. To
address this problem we introduced a simple eligibility trace by
evaluating the activity of the pre- and post synaptic units in the
learning rule with a small delay $\delta t$ (see supplementary material). With this addition, the network model is
able to find the optimal solution for this task within roughly 2000
steps (\fref{fig:grid_world}{\bf A}). It also learns to associate
holes with punishment and frozen states with reward if they are on the
path to the goal (\fref{fig:grid_world}{\bf B}). Although there are
two possible paths to the goal, the agent prefers the path with less
corners, since it is easier to navigate using a WTA circuit.

\section{Conclusion}
In this manuscript, we have presented a toolchain that closes the loop between the OpenAI Gym and neural network simulators.
We demonstrated the functionality of the toolchain by implementing an actor-critic architecture in NEST and evaluating its performance on two different environments.
The performance of the network quickly reached near-optimal performance on these two tasks.

Combining neural network simulators with RL toolkits responds to the growing need of researchers to provide neural network models with rich, dynamic input.
Compared to creating customized environments to this end, using readily available tools is easier, often computationally more efficient, and most importantly, supports reproducible science.
In addition, having the OpenAI Gym environments as common benchmarks in both fields encourages comparison between traditional machine learning and biologically plausible implementations.
In contrast to models presented in previous studies, our toolchain makes it easy for other researchers to extend our implementation of an actor-critic architecture to other environments, replace neurons models or explore alternative learning rules.

While the toolchain currently only supports the OpenAI Gym, the extension to other toolkits is simple due to a modular design of the wrapper.
The RMT can be found on GitHub and is available under the GPLv3. 
The OpenAI Gym ZeroMQ wrapper is also available via GitHub under the MIT license. 
A complementary development to the work presented here is provided by SPORE, a framework for reward-based learning with spiking neurons in the NEST simulator.\footnote{\href{https://github.com/IGITUGraz/spore-nest-module}{https://github.com/IGITUGraz/spore-nest-module}}
It provides support for synapse models with time-driven updates, additional support for recording and evaluating traces of neuronal state variables and introduces MUSIC ports for communicating rewards to a running simulation.

With the work presented here we enable researchers to build more easily upon previous studies and evaluate novel models.
We hope this boosts the progress in computational neuroscience in uncovering the biophysical mechanisms involved in learning complex tasks from delayed rewards.


\subsubsection*{Acknowledgments}
We acknowledge partial support by the German Federal Ministry of Education through our German-Japanese Computational Neuroscience Project (BMBF Grant 01GQ1343), EuroSPIN, the Helmholtz Alliance through the Initiative and Networking Fund of the Helmholtz Association and the Helmholtz Portfolio theme ``Supercomputing and Modeling for the Human Brain'' and the European Union Seventh Framework Programme (FP7/2007-2013) under grant agreement no. 604102 (HBP). All network simulations carried out with NEST (http://www.nest-simulator.org).


\appendix

\section*{Supplementary material}
\section{Derivation of the learning rule}
We consider continuous time, continuous states and continuous actions and follow similar steps as \cite{Doya00,Fremaux2013_e1003024}. Starting from the continuous-time value function
\begin{align}
  V^\pi(\vec{s}(t)):= \int_t^\infty r(\vec{s}^\pi(t'))e^{-\frac{t'-t}{\tau_r}}dt' \, ,
\end{align}
we take the derivative with respect to $t$ to arrive at a self-consistency equation for $V$:
\begin{align}
  \dot{V}^\pi(t) := \frac{dV^\pi(t)}{dt} = -r(t) + \frac{1}{\tau_r} V^\pi(t) \, .
  \label{eq:self-consistency}
\end{align}
To implement temporal-difference learning in a neural network architecture, we would like to approximately represent the true value $V^\pi(t)$ of the state at time $t$ by the rate $z_i(t)$ of a critic neuron. This activity will depend on the activity of input units, i.e., place cells, and the weights between inputs and critic.
With initially random weights the self-consistency criterion will not be fulfilled, and will have a finite error $\delta(t)$:
\begin{align}
  \delta(t) = \dot{z}_i(t) + r(t) - \frac{1}{\tau_r} z_i(t) \, .
\end{align}
We now define an objective function that should be minimized by gradient descent on the weights:
\begin{align}
  E(t) = \frac{1}{2}(V^\pi(t) - z_i(t))^2 \, .
\end{align}
We take derivative with respect to $w_{ij}$ and use the self-consistency \fref{eq:self-consistency}:
\begin{align}
  \frac{\partial E(t)}{\partial w_{ij}} &= -(V^\pi(t) - z_i(t))\frac{\partial z_i(t)}{\partial w_{ij}} \notag \\
                                        &= -(\tau_r \dot{V}^\pi(t) + \tau_r r(t) - z_i(t))\frac{\partial z_i(t)}{\partial w_{ij}} \notag \\
                                        &\approx -\underbrace{(\tau_r \dot{z}(t) + \tau_r r(t) - z_i(t))}_{\tau_r\delta(t)}\frac{\partial z_i(t)}{\partial w_{ij}}
\end{align}
Here we have replaced $\dot{V}^\pi(t)$ with $\dot{z}_i(t)$ in the last line. For a discussion of the validity of this approximation and the convergence of the learning rule, see \cite{Fremaux2013_e1003024}.
To perform gradient descent on the objective function, we hence need to change the weights according to 
\begin{align}
  \Delta w_{ij} :=& -\eta' \frac{\partial E(t)}{\partial w_{ij}} \notag \\
  =& \eta \delta(t) \frac{\partial z_i(t)}{\partial w_{ij}} \, ,
  \label{eq:deltaw-with-derivative}
\end{align}
where we introduced $\eta=\eta'\tau_r$.
We hence need to determine the derivative of the critic activity with respect to the weights between inputs and critic $\frac{\partial z_i(t)}{\partial w_{ij}}$.

We start from the differential equation describing the dynamics of a threshold-linear rate neuron, and assume the noise to be small, i.e., we drop the term $\xi_i(t)$. Without loss of generality, we assume $\mu_i=0, \theta_i=0$. The dynamics are then given by the solution to
\begin{align}
  \tau \frac{dz_i(t)}{dt}=-z_i(t) + g\, \Theta\left( \sum_j w_{ij}x_j(t) \right)\left( \sum_j w_{ij}x_j(t) \right) \, .
\end{align}
Variation of constants yields the general solution as a convolution equation:
\begin{align}
  z_i(t) = g \left(\left(\Theta\left(\dots \right)\left( \sum_j w_{ij}x_j(\cdot) \right)\right) \ast \kappa(\cdot) \right)(t) \, ,
\end{align}
where we have introduced the filter kernel $\kappa(t) = \frac{1}{\tau} e^{-\frac{t}{\tau}} \Theta(t)$.
We now take the derivative with respect to $w_{ij}$. While the derivative at $\sum_j w_{ij}x_j(t) = 0$ is technically not defined, we follow the usual convention and set it to zero. This yields
\begin{align}
  \frac{\partial z_i(t)}{\partial w_{ij}} =
  \begin{cases}
    g\left(x_j \ast \kappa\right)(t) & \text{if}\, \sum_j w_{ij}x_j(t) > 0 \\
    0 & \text{else}
  \end{cases}
  \label{eq:derivative-critic}
\end{align}
By combining \fref{eq:deltaw-with-derivative} with \fref{eq:derivative-critic}, we obtain the following learning rule:
\begin{align}
  \Delta w_{ij} =
  \begin{cases}
    \eta \delta(t) g (x_j \ast \kappa)(t) & \text{if}\, \sum_j w_{ij}x_j(t) > 0 \\
    0 & \text{else}
  \end{cases}
\end{align}
By choosing the time constant of critic and actors small, we effectively remove the filtering of the presynaptic activity ($\lim_{\tau \rightarrow 0}\kappa(t) = \delta(t)$) and hence ignore it. To simplify this equation further, we rewrite it as a condition on the rate of the postsynaptic neuron by observing that $z_i(t) > 0\, \text{iff} \sum_j w_{ij}x_j(t) > 0$. To implement exploration for similar inputs to all output units we add noise to the activity of the actor units. We only consider a postsynaptic neuron active, if its activity is larger than some threshold $\theta_\text{post}$. This leads to the following form for the learning rule:
\begin{align}
  \Delta w_{ij} = \eta \delta(t) g x_j(t) \Theta(z_i(t) - \theta_\text{post}) \, ,
\end{align}
where we $\Theta(\cdot)$ denotes the Heaviside step function defined as:
\begin{align}
  \notag
  \Theta(x) =
  \begin{cases}
    1 & x > 0 \\
    0 & \text{else}
  \end{cases}
\end{align}

To implement a simple type of eligibility trace, we introduce an additional parameter $\delta t$ that can delay the activity of the pre- and post-synaptic units in the learning rule:
\begin{align}
  \Delta w_{ij} = \eta \delta(t) g x_j(t - \delta t) \Theta(z_i(t - \delta t) - \theta_\text{post}) \, ,
\end{align}


\section{Network description}

The tables \ref{tab:nordlie} and \ref{tab:params} summarize the network architecture and parameters.

\newlength{\columnfigwidth}
\newlength{\fullfigwidth}
\setlength{\columnfigwidth}{8.6cm} 
\setlength{\fullfigwidth}{17.2cm}

\newlength{\columnwidthleft}
\newlength{\columnwidthmiddle}

\newcommand{\modelhdr}[3]{
  \multicolumn{#1}{|l|}{
    \color{white}
    \cellcolor[gray]{0.0}
    \textbf{\makebox[0pt][l]{#2}\hspace{0.5\fullfigwidth}\makebox[0pt][c]{#3}}
  }
}
\newcommand{\parameterhdr}[3]{
  \multicolumn{#1}{|l|}{
    \color{black}\cellcolor[gray]{0.8}
    \textbf{\makebox[0pt][l]{#2}\hspace{0.5\fullfigwidth}\makebox[0pt][c]{#3}}
  }
}

\begin{table*}

\setlength{\columnwidthleft}{0.2\textwidth}
\setlength{\columnwidthmiddle}{0.2\textwidth}

\begin{tabularx}{\fullfigwidth}{|p{\columnwidthleft}|X|}
  \hline\modelhdr{2}{A}{Model summary}\\\hline
  Populations    & Seven \\
  \hline
  Topology       & None \\
  \hline
  Connectivity   & Population specific \\
  \hline
  Neuron model   & Linear \& threshold-linear rate \\
  \hline
  Channel models & None \\
  \hline
  Synapse model  & Instantaneous \& delayed continuous coupling \\
  \hline
  Plasticity     & Three-factor Hebbian \\
  \hline
  External input & Continuous MUSIC ports \\
  \hline
  External output & Continuous MUSIC ports \\
  \hline
  Measurements   & Rates of all neurons \\
  \hline
\end{tabularx} \\

\begin{tabularx}{\fullfigwidth}{|p{\columnwidthleft}|p{\columnwidthmiddle}|X|}
  \hline\modelhdr{3}{B}{Populations}\\\hline
  \bf Name & \bf Elements & \bf Size \\
  \hline
  Observation & MUSIC in port & 1 \\
  Reward & MUSIC in port & 1 \\
  Action & MUSIC out port & 1 \\
  Place cells & Threshold-linear & $16 (25)$ \\
  Critic & Threshold-linear & $1$ \\
  Actor & Threshold-linear & $4 (3)$ \\
  Prediction error & Linear & $1$ \\
  \hline
\end{tabularx} \\

\begin{tabularx}{\fullfigwidth}{|p{\columnwidthleft}|p{\columnwidthmiddle}|X|}
  \hline\modelhdr{3}{C}{Connectivity}\\\hline
  \bf Source & \bf Target & \bf Pattern \\
  \hline
  Observation & Place cells & One-to-one (by MUSIC channel), instantaneous, static, weight $w_\text{o}$ \\
  Reward & Prediction error & One-to-one (by MUSIC channel), instantaneous, static, weight $w_\text{r}$ \\
  Actor & Action & One-to-one (by MUSIC channel), instantaneous, static, weight $w_\text{a}$ \\
  Place cells & Critic & All-to-all, instantaneous, plastic, initial weight $w_\text{pc}$ \\
  Place cells & Actor & All-to-all, instantaneous, plastic, initial weight $w_\text{pa}$ \\
  Critic & Prediction error & One-to-one, instantaneous, static, weight $1/d - 1/\tau_r$ \\
  Critic & Prediction error & One-to-one, delay $d$, static, weight $-1/d$ \\
  Actor & Actor & All-to-all, instantaneous, static, weight $\alpha \exp(-\Delta a / \sigma) + \beta$ \\
  \hline
\end{tabularx} \\

\begin{tabularx}{\fullfigwidth}{|p{\columnwidthleft}|X|}
  \hline\modelhdr{2}{D}{Neuron and synapse model}\\\hline
  Type & Linear rate neuron \\
  Dynamics & $\tau \frac{dz(t)}{dt}=-z(t) + \mu + \left(h(t) - \theta\right) + \xi(t)$ \\
  \hline
  Type & Threshold-linear rate neuron \\
  Dynamics & $\tau \frac{dz(t)}{dt}=-z(t) + \mu + \Theta\left(h(t) - \theta\right)\left(h(t) - \theta\right) + \xi(t)$ \\
  \hline
  Type & Three-factor Hebbian synapse \\
  Plasticity & $\Delta w_{ij} = \eta \delta(t) g x_j(t - \delta t) \Theta(z_i(t - \delta t) - \theta_\text{post})$ \\
  \hline
\end{tabularx}

\begin{tabularx}{\fullfigwidth}{|p{\columnwidthleft}|X|}
  \hline\modelhdr{2}{E}{Input}\\\hline
  \bf Type & \bf Description \\
  \hline
  Observation & Rate $r\in [-1, 1]$ according to tuning of place cell (using {\it discretize} adapter) \\
  \hline
  Reward & Rate $r\in [-1, 1]$ according to reward provided by the environment \\
  \hline
\end{tabularx}

\begin{tabularx}{\fullfigwidth}{|p{\columnwidthleft}|X|}
  \hline\modelhdr{2}{F}{Output}\\\hline
  \bf Type & \bf Description \\
  \hline
  Action & Rates $r_i \in [0, \infty)$ according to activities of the actor units \\
  \hline
  \hline
\end{tabularx}

\caption{
  Description of the network model (according to \citep{Nordlie-2009_e1000456}).
  \label{tab:nordlie}
}

\end{table*}

\begin{table*}

\setlength{\columnwidthleft}{0.2\textwidth}
\setlength{\columnwidthmiddle}{0.2\textwidth}

\begin{tabularx}{\fullfigwidth}{|p{\columnwidthleft}|X|}
  \hline\parameterhdr{2}{B}{Populations: place cells}\\\hline
  \bf Name & \bf Values \\
  \hline
  $\tau$ & 5.0 (1.0) \\
  \hline
  $g$ & 1.0 \\
  \hline
  $\mu$ & 0.0 \\
  \hline
  $\sigma_\xi$ & 0.0 \\
  \hline
  $\theta$ & -0.5 \\
  \hline
  \hline\parameterhdr{2}{B}{Populations: critic}\\\hline
  \bf Name & \bf Values \\
  \hline
  $\tau$ & 0.1 \\
  \hline
  $g$ & 1.0 \\
  \hline
  $\mu$ & -1.0 \\
  \hline
  $\sigma_\xi$ & 0.0 \\
  \hline
  $\theta$ & -1.0 \\
  \hline
  \hline\parameterhdr{2}{B}{Populations: reward}\\\hline
  \bf Name & \bf Values \\
  \hline
  $\tau$ & 1.0 \\
  \hline
  $g$ & 1.0 \\
  \hline
  $\mu$ & 0.0 \\
  \hline
  $\sigma_\xi$ & 0.0 \\
  \hline
  $\theta$ & 0.001 (-0.0999) \\
  \hline
  \hline\parameterhdr{2}{B}{Populations: actor}\\\hline
  \bf Name & \bf Values \\
  \hline
  $\tau$ & 0.1 \\
  \hline
  $g$ & 1.0 \\
  \hline
  $\mu$ & 0.0 \\
  \hline
  $\sigma_\xi$ & 0.2 (0.05) \\
  \hline
  $\theta$ & 0.0 \\
  \hline
  \hline\parameterhdr{2}{C}{Connectivity}\\\hline
  \bf Name & \bf Values \\
  \hline
  $w_\text{o}$ & $0.5$ \\
  \hline
  $w_\text{r}$ & $0.1$ \\
  \hline
  $w_\text{a}$ & $1.0$ \\
  \hline
  $w_\text{pc}$ & $0.0 $ \\
  \hline
  $w^\text{min}_\text{pc}$ & $-1.0$ \\
  \hline
  $w^\text{max}_\text{pc}$ & $1.0$ \\
  \hline
  $\theta_\text{post}^\text{pc} $ & $-1.0$ \\  
  \hline
  $w_\text{pa}$ & $0.9 (0.3)$ \\
  \hline
  $w^\text{min}_\text{pa}$ & $0.1 (0.05)$ \\
  \hline
  $w^\text{max}_\text{pa}$ & $1.0$ \\
  \hline
  $\theta_\text{post}^\text{pa} $ &  $0.5 (0.1)$ \\
  \hline
  $d$ & $1.0$ \\
  \hline
  $\tau_r$ & $20000.0$ \\
  \hline
  $\alpha$ & $1.2$ \\
  \hline
  $\beta$ & $-0.55$ \\
  \hline
  $\sigma_\xi$ & $0.1$ \\
  \hline
  $\eta_\text{critic}$ & $0.01 (0.125)$ \\
  \hline
  $\eta_\text{actor}$ & $0.2 (0.250)$ \\
  \hline
  $\delta t$ & $19.0 (0.0)$ \\
  \hline\parameterhdr{2}{E}{Input: discretize adapter}\\\hline
  \bf Name & \bf Values \\
  \hline
  $\sigma_x$ & 0.01 (0.2) \\
  $\sigma_y$ & -- (0.2) \\
 \hline
\end{tabularx} \\

\caption{
    Table of the network parameters used for both tasks (according to \citep{Nordlie-2009_e1000456}). Values in brackets are used for the \emph{MountainCar} environment.
  \label{tab:params}
}

\end{table*}

\section{JSON Message types}

Listing \ref{lst:message_types} show the standard message types used for communication between the OpenAI Gym and the RMT. All messages are serialized using JSON and communicated via ZeroMQ.

\begin{lstlisting}[caption=Message types used for communication.,label=lst:message_types]
BasicMsg
{
    float min
    float max
    float value
    float timestamp
}

ObservationMsg
{
    BasicMsg[] observations   # one basic msg per dimension
}

RewardMsg
{
    BasicMsg[] reward         # reward is always one dimensional
}

ActionMsg
{
    BasicMsg[] actions        # one dimensional for discrete actions
                              # or one dimension per possible action
}
\end{lstlisting}

\section{Example wrapper configuration file}

Listing \ref{lst:zmq-wrapper-conf} shows an example configuration file for running the mountain car environment.
\begin{lstlisting}[caption=Example configuration file for the wrapper to run the ``MountainCar-v0'' environment.,label=lst:zmq-wrapper-conf]
    "All":
    {
        "seed": 12345,
        "time_stamp_tolerance": 0.01,
        "prefix": null,
        "write_report": true,
        "report_file": "./report.json",
        "overwrite_files": false,
        "flush_report_interval": null
    },
    "Env":
    {
        "env": "MountainCar-v0",
        "initial_reward": null,
        "final_reward": null,
        "min_reward": -1.0,
        "max_reward": 1.0,
        "render": true,
        "monitor": false,
        "monitor_dir": "./experiment-0/",
        "monitor_args":
        {
            "write_upon_reset": true,
            "video_callable": false
        }
    },
    "EnvRunner":
    {
        "update_interval": 0.01,
        "inter_trial_duration": 0.4
    },
    "CommandReceiver":
    {
        "socket": 5555,
        "time_stamp_tolerance": 0.01
    },
    "ObservationSender":
    {
        "socket": 5556,
        "update_interval": 0.01
    },
    "RewardSender":
    {
        "socket": 5557,
        "update_interval": 0.01
    }
\end{lstlisting}

\section{Example MUSIC configuration file}
Listing \ref{lst:music-conf} shows an example MUSIC configuration file to run the MountainCar environment. It shows the different processes with parameters which are spawned by MUSIC including RMT adapters and NEST. 
\begin{lstlisting}[caption=Example MUSIC configuration file to run the MountainCar environment.,label=lst:music-conf]
stoptime=150.
rtf=1.
[reward]
  binary=zmq_in_adapter
  args=
  np=1
  music_timestep=0.001
  message_type=GymObservation
  zmq_topic=
  zmq_addr=tcp://localhost:5557
[sensor]
  binary=zmq_in_adapter
  args=
  np=1
  music_timestep=0.001
  message_type=GymObservation
  zmq_topic=
  zmq_addr=tcp://localhost:5556
[discretize]
  binary=discretize_adapter
  args=
  np=1
  music_timestep=0.001
  grid_positions_filename=grid_pos.json
[nest]
  binary=../actor_critic_network/network.py
  args=-t 150. -n 25 -m 3 -p network_params.json
  np=1
[argmax]
  binary=argmax_adapter
  args=
  np=1
  music_timestep=0.001
[command]
  binary=zmq_out_adapter
  args=
  np=1
  music_timestep=0.01
  message_type=GymCommand
  zmq_topic=
  zmq_addr=tcp://*:5555
sensor.out->discretize.in[2]
discretize.out->nest.in[25]
reward.out->nest.reward_in[1]
nest.out->argmax.in[3]
argmax.out->command.in[1]
\end{lstlisting}

%
%

\section{Environments}

Table \ref{tab:env} shows parameters for the OpenAI Gym environments and the ZeroMQ wrapper.

\begin{table*}
    \label{tab:env}

\setlength{\columnwidthleft}{0.4\textwidth}
\setlength{\columnwidthmiddle}{0.2\textwidth}

\begin{tabularx}{\fullfigwidth}{|p{\columnwidthleft}|X|}
  \hline\parameterhdr{2}{ }{OpenAI Gym}\\\hline
  \bf Name & \bf Values \\
    \hline
    Version & 0.8.1 \\
  \hline\parameterhdr{2}{ }{MountainCar}\\\hline
  \bf Name & \bf Values \\
    \hline
    Version & 0 \\
    \hline
    Max episode steps & None \\
    \hline
    Initial reward* & -1.0 \\
    \hline
    Final reward* & -0.4 \\
    \hline
    Inter-trial duration* & 0.4 \\
    \hline
    Update interval (env runner)* & 0.02 \\
  \hline\parameterhdr{2}{ }{FrozenLake}\\\hline
  \bf Name & \bf Values \\
    \hline
    Version & 0 \\
    \hline
    Max episode steps & None \\
    \hline
    Slippery & False \\
    \hline
    Final reward null* & -0.1 \\
    \hline
    Inter-trial duration* & 0.1 \\
    \hline
    Update interval (env runner)*& 0.1 \\
 \hline
\end{tabularx} \\

\caption{Table of the environment parameters. Values marked with * indicate values for the ZeroMQ wrapper.}

\end{table*}


\end{document}